\documentclass[11pt]{article}
\usepackage{graphicx}
\usepackage{amsmath}
\usepackage{amssymb}
\usepackage{caption2}
\begin{document}
\bibliographystyle{prsty}
\begin{center}
{\large {\bf \sc{  Analysis of  the triply heavy  baryon states   with QCD sum rules }}} \\[2mm]
Zhi-Gang Wang \footnote{E-mail:wangzgyiti@yahoo.com.cn.  }     \\
 Department of Physics, North China Electric Power University,
Baoding 071003, P. R. China
\end{center}

\begin{abstract}
In this article, we  study the  ${1\over 2}^{\pm}$ and ${3\over 2}^{\pm}$ triply heavy
baryon states in a systematic way   by subtracting the contributions from the corresponding ${1\over 2}^{\mp}$ and ${3\over 2}^{\mp}$ triply heavy
baryon states with the QCD sum rules, and make reasonable predictions for their masses.
\end{abstract}



\section{Introduction}

By this time, the ${1\over 2}^+$ and ${1\over 2}^-$ antitriplet charmed
baryon states ($\Lambda_c^+$, $\Xi_c^+$, $\Xi_c^0)$ and
($\Lambda_c^+(2595)$, $\Xi_c^+(2790)$, $\Xi_c^0(2790))$,  and the
${1\over 2}^+$ and ${3\over 2}^+$ sextet charmed baryon states
($\Omega_c,\Sigma_c,\Xi'_c$) and ($\Omega_c^*,\Sigma_c^*,\Xi^*_c$)
have been observed, while the $S$-wave bottom baryon
states are far from complete, only the $\Lambda_b^0$, $\Sigma_b^+$, $\Sigma_b^0$, $\Sigma_b^-$,
$\Sigma_b^{*+}$, $\Sigma_b^{*-}$, $\Xi_b^0$, $\Xi_b^-$ and $\Omega_b^-$ have been observed \cite{PDG}.
In 2002, the SELEX collaboration reported the first observation of  the doubly charmed baryon state
$ \Xi_{cc}^+$ in the
 decay  $\Xi_{cc}^+\rightarrow\Lambda_c^+K^-\pi^+$  \cite{SELEX2002}, and
confirmed it later   in the decay
$\Xi_{cc}^+\rightarrow pD^+K^- $   \cite{SELEX2004}. However, the Babar collaboration  observed no evidence for the $\Xi_{cc}^+$ in the
  $\Lambda_c^+K^-\pi^+$, $\Xi_c^0\pi^+$ decay modes and for the $\Xi_{cc}^{++}$ in the
  $\Lambda_c^+K^-\pi^+\pi^+$, $\Xi_c^0\pi^+\pi^+$ decay modes, and the
Belle collaboration    observed no evidence  for the $\Xi_{cc}^+$ in the   $\Lambda_c^+K^-\pi^+$ decay mode
\cite{Xi-cc-Babar,Xi-cc-Belle}. There are no experimental signals  for the  triply heavy baryon states, we
expect that the large hadron collider (LHC) will provide
us with the whole heavy, doubly heavy and triply heavy baryon states \cite{LHC,LHC-prod}.

The triply heavy baryon states and heavy quarkonium states  play an important role  in understanding the heavy quark dynamics  at the
hadronic scale due to the absence  of the light quark contaminations, and serve as an excellent subject
  in studying   the interplay between the perturbative and nonperturbative QCD. On the other hand,
the  QCD sum rules is a powerful nonperturbative theoretical tool in studying the
ground state hadrons \cite{SVZ79,NarisonBook}. In the
QCD sum rules, the operator product expansion is used to expand the
time-ordered currents into a series of quark and gluon condensates
which parameterize the long distance properties. Taking
the quark-hadron duality, we can obtain copious information about
the hadronic parameters at the phenomenological side
\cite{SVZ79,NarisonBook}. There have been many works on the
masses of the heavy and doubly heavy baryon states with the  QCD sum rules  \cite{H-baryon,H-baryon-Wang}.
It is interesting to study  the mass spectrum  of the triply heavy   baryon states using  the QCD sum rules.

In Ref.\cite{H-baryon-Wang}, we take the novel  approach introduced by Jido et al \cite{Oka96} to
study the  positive-parity and negative-parity heavy and doubly heavy baryons in a systematic way
by separating the contributions of  the positive-parity and negative-parity baryons explicitly,
as the interpolating currents  have non-vanishing
couplings to both the positive-parity and negative-parity baryons, there exist  contaminations
\cite{Chung82}. Before the work of Jido et al,  Bagan et al take the heavy quark limit  to separate the
contributions of the positive-parity and negative-parity heavy baryons
 unambiguously  \cite{Bagan93}. In this article,  we
extend our previous works to study the ${1\over 2}^{\pm}$ and ${3\over
2}^{\pm}$ triply heavy baryon states
by subtracting the contributions from the corresponding
 ${1\over 2}^{\mp}$ and ${3\over
2}^{\mp}$ triply heavy baryon states with the full QCD sum rules.

The existing theoretical works focus on  the heavy and doubly heavy baryon states, the works on
the triply heavy baryon states  are relatively
few, for example, the effective field theory \cite{EFT}, the lattice QCD
\cite{LQCD}, the  QCD bag model \cite{HaseM}, the quark model estimation
\cite{BjorM},  the variational approach \cite{JiaM}, the modified bag model
\cite{BeroM}, the relativistic three-quark model \cite{MartM}, the QCD sum rules \cite{ZhanM},
the non-relativistic three-quark model \cite{RobeM,Vijande2004,Patel2009},  potential non-relativistic QCD \cite{Llanes-Estrada},
the Regge trajectory ansatz \cite{XHGuo}, etc.

 The article is arranged as follows:  we derive the
QCD sum rules for the masses and the pole residues of  the triply
 heavy baryon states in Sect.2;
 in Sect.3, we present the numerical results and discussions; and Sect.4 is reserved for our
conclusions.

\section{QCD sum rules for  the triply  heavy baryon states }
The ${1\over 2}^+$ and ${3\over 2}^+$ triply heavy baryon states
$\Omega_{QQQ'}({1\over 2})$, $\Omega_{QQQ'}({3\over 2})$  and
$\Omega_{QQQ}({3\over 2})$ can be interpolated by the triply heavy quark currents
$J^{QQQ'}(x)$, $J^{QQQ'}_\mu(x)$  and $J^{QQQ}_\mu(x)$,  respectively,
\begin{eqnarray}
J^{QQQ'}(x)&=& \epsilon^{ijk}  Q^T_i(x)C\gamma_\mu Q_j(x) \gamma^\mu \gamma_5   Q'_k(x) \, , \nonumber \\
J^{QQQ'}_\mu(x)&=& \epsilon^{ijk}  Q^T_i(x)C\gamma_\mu Q_j(x)    Q'_k(x) \, , \nonumber \\
J^{QQQ}_\mu(x)&=& \epsilon^{ijk}  Q^T_i(x)C\gamma_\mu Q_j(x)    Q_k(x) \, ,
\end{eqnarray}
where the  $Q$ and $Q'$ represent the heavy quarks $c$ and $b$,  the $i$,
$i$ and $k$ are color indexes, and the $C$ is the charge conjunction
matrix. There are other currents such as the  $\eta^{QQQ'}$ besides the Ioffe currents $J^{QQQ'}$ to interpolate the ${\frac{1}{2}}^+$ triply heavy baryon states,
\begin{eqnarray}
\eta^{QQQ'}(x)&=& \epsilon^{ijk}  Q^T_i(x)C\sigma_{\mu\nu} Q_j(x) \sigma^{\mu\nu} \gamma_5   Q'_k(x) \, .
\end{eqnarray}
The currents $J^{QQQ'}$ and $\eta^{QQQ'}$ correspond to the superimpositions  $ \mathcal{O}^{QQQ'}_1- \mathcal{O}^{QQQ'}_2$ and
$ \mathcal{O}^{QQQ'}_1+ \mathcal{O}^{QQQ'}_2$ respectively,
 where  the fundamental currents $ \mathcal{O}^{QQQ'}_1$ and $ \mathcal{O}^{QQQ'}_2$ are defined as
\begin{eqnarray}
\mathcal{O}^{QQQ'}_1(x)&=&\epsilon^{ijk}Q^T_i(x)C Q'_j(x) \gamma_5 Q_k(x)\, ,\nonumber \\
\mathcal{O}^{QQQ'}_2(x)&=&\epsilon^{ijk}Q^T_i(x)C\gamma_5 Q'_j(x) Q_k(x)\, .
\end{eqnarray}
We can take the simple replacements $Q \to u$ and $Q' \to d$ to obtain the corresponding currents for the proton \cite{Chung82,CDKS}.
The convergent behavior of the current $\eta^{uud}$ is not as good as the Ioffe current $J^{uud}$, and
 appearance of  chirality violation terms in the operator product  expansion indicates that the current $\eta^{uud}$ couples strongly both to the positive-parity and negative-parity baryon states  \cite{Ioffe1983}.   We can also choose the most general current $\mathcal{O}^{uud}$,
 \begin{eqnarray}
 \mathcal{O}^{uud}(x)&=&\mathcal{O}^{uud}_1(x)+t\mathcal{O}^{uud}_2(x)\, ,
 \end{eqnarray}
and search for the ideal value $t$. Detailed studies of all the octet baryon states based on the QCD sum rules indicate that the optimal value is about $t=-1$,
 if the experimental values of the masses are taken as the input parameters \cite{Tarrach}, i.e. the Ioffe currents work well. We expect that the conclusion survives in the case of the heavy quark systems. At the present time,  no experimental data are available   to be taken as the input parameters in searching for the optimal value of the $t$.

 The corresponding ${1\over 2}^-$ and ${3\over 2}^-$ triply heavy baryon states can be
interpolated by the  currents $J^{-} =i\gamma_{5} J^{+}$ and $J^{-}_\mu =i\gamma_{5} J^{+}_{\mu}$
because multiplying $i \gamma_{5}$ to the currents $J^{+}$  and $J^{+}_\mu$ changes their
parity \cite{Oka96}, where the currents $J^{+}$ and $J^{+}_\mu$
denotes the triply heavy quark currents $J^{QQQ'}(x)$, $J^{QQQ'}_\mu(x)$  and $J^{QQQ}_\mu(x)$, respectively.

The correlation functions $\Pi^{\pm}(p)$ and $\Pi^{\pm}_{\mu\nu}(p)$ are defined by
\begin{eqnarray}
\Pi^{\pm}(p)&=&i\int d^4x e^{ip \cdot x} \langle0|T\left\{J^{\pm}(x)\bar{J}^{\pm}(0)\right\}|0\rangle \, , \nonumber \\
\Pi^{\pm}_{\mu\nu}(p)&=&i\int d^4x e^{ip \cdot x} \langle0|T\left\{J^{\pm}_\mu(x)\bar{J}^{\pm}_{\nu}(0)\right\}|0\rangle \, .
\end{eqnarray}
The currents $J^{\pm}(x)$ couple  to the ${\frac{1}{2}}^{\pm}$ triply heavy baryon
states $B_{\pm}$, while the currents $J_{\mu}^{\pm}(x)$ couple  to
both the ${\frac{3}{2}}^{\pm}$ triply heavy  baryon states $B^*_{\pm}$ and the
${\frac{1}{2}}^{\pm}$ triply heavy baryon states $B_{\pm}$ \cite{Chung82},
\begin{eqnarray}
   \langle{0}|J^{+}(0)| B_{\pm}(p)\rangle \langle B_{\pm}(p)|\bar{J}^{+}(0)|0\rangle &=& - \gamma_{5}\langle 0|J^{-}(0)| B_{\pm}(p)\rangle \langle B_{\pm}(p)| \bar{J}^{-}(0)|0\rangle \gamma_{5} \, , \nonumber \\
\langle{0}|J^{+}_{\mu}(0)| B_{\pm}^*(p)\rangle \langle B_{\pm}^*(p)|\bar{J}^{+}_{\nu}(0)|0\rangle &=&
   - \gamma_{5}\langle 0|J^{-}_{\mu}(0)| B_{\pm}^*(p)\rangle \langle B_{\pm}^*(p)| \bar{J}^{-}_{\nu}(0)|0\rangle \gamma_{5} \,, \nonumber \\
    \langle{0}|J^{+}_{\mu}(0)| B_{\pm}(p)\rangle \langle B_{\pm}(p)|\bar{J}^{+}_{\nu}(0)|0\rangle
&=&    - \gamma_{5}\langle 0|J^{-}_{\mu}(0)| B_{\pm}(p)\rangle
\langle B_{\pm}(p)| \bar{J}^{-}_{\nu}(0)|0\rangle \gamma_{5} \, ,
\end{eqnarray}
where
\begin{eqnarray}
\langle 0| J^{\pm} (0)|B_{\pm}(p)\rangle &=&\lambda_{\pm} U(p,s) \, , \nonumber \\
\langle 0| J^{\pm}_\mu (0)|B_{\pm}^*(p)\rangle &=&\lambda_{\pm} U_\mu(p,s) \, , \nonumber \\
 \langle0|J^{\pm}_{\mu}(0)|B_{\mp}(p)\rangle&=&\lambda_{\mp}
 \left(\gamma_{\mu}-4\frac{p_{\mu}}{M_{\mp}}\right)U(p,s) \, ,
\end{eqnarray}
the $\lambda_{\pm}$  are  the  pole residues, the $M_{\pm}$ are
the masses, and  the Dirac spinors $U(p,s)$ and $U_\mu(p,s)$  satisfy
the following identifies,
\begin{eqnarray}
\sum_sU(p,s)\overline {U}(p,s)&=&\!\not\!{p}+M_{\pm} \, , \nonumber \\
\sum_s U_\mu(p,s) \overline{U}_\nu(p,s)
&=&(\!\not\!{p}+M_{\pm})\left[ -g_{\mu\nu}+\frac{\gamma_\mu\gamma_\nu}{3}+\frac{2p_\mu p_\nu}{3M_{\pm}^2}-\frac{p_\mu\gamma_\nu-p_\nu \gamma_\mu}{3M_{\pm}} \right] \, . \end{eqnarray}

We  insert  a complete set  of intermediate triply heavy baryon states with the same quantum numbers as the current operators $J^{\pm}(x)$ and
$J_\mu^{\pm}(x)$ into the correlation functions $\Pi^{\pm}(p)$ and
$\Pi^{\pm}_{\mu\nu}(p)$ to obtain the hadronic representation
\cite{SVZ79}. After isolating the pole terms of the lowest
states of the positive-parity and negative-parity  triply heavy baryons, we obtain the
following results  \cite{Oka96}:
\begin{eqnarray}
 \Pi^{\pm}(p) & = &\lambda_+^2 {\!\not\!{p} + M_{+} \over M^{2}_+ -p^{2} } + \lambda_{-}^2 {\!\not\!{p} - M_{-} \over M_{-}^{2}-p^{2} } + \cdots \,, \nonumber \\
 \Pi^{\pm}_{\mu\nu}(p)&=&-\Pi_{\pm}(p)g_{\mu\nu}+\cdots \,, \nonumber \\
\Pi_{\pm}(p) &=&\lambda_+^2 {\!\not\!{p}+M_{+}\over M^{2}_{+}-p^{2}}+\lambda_{-}^2{\!\not\!{p}-M_{-}\over M_{-}^{2}-p^{2}}+\cdots\, .
    \end{eqnarray}
 In this article, we choose the tensor structure $g_{\mu\nu}$ for analysis, the ${1\over
2}^\pm$ triply heavy baryon states have no contaminations.

 We can take $\vec{p} = 0$ for the correlation functions $\Pi(p)$ ($\Pi^{\pm}(p)$, $\Pi_{\pm}(p)$), and obtain the spectral densities  at the phenomenological side,
\begin{eqnarray}
  \rm{limit}_{\epsilon\rightarrow0}\frac{{\rm Im}  \Pi(p_{0}+i\epsilon)}{\pi} & = & \lambda_+^2 {\gamma_{0} + 1\over 2} \delta(p_{0} - M_+)+
    \lambda_{-}^{2} {\gamma_{0} - 1\over 2} \delta(p_{0} - M_{-}) +\cdots \nonumber \\
  & = & \gamma_{0} A(p_{0}) + B(p_{0})+\cdots \, ,
\end{eqnarray}
where
\begin{eqnarray}
  A(p_{0}) & = &  {1 \over 2}\left[\lambda_+^{2}  \delta(p_{0} - M_+)+\lambda_-^{2} \delta(p_{0} -  M_{-})  \right]  \, , \nonumber \\
  B(p_{0}) & = & {1 \over 2} \left[ \lambda_+^{2}  \delta(p_{0} - M_+)  - \lambda_-^{2} \delta(p_{0} -  M_{-})\right] \, ,
\end{eqnarray}
the combinations  $A(p_{0}) + B(p_{0})$ and $A(p_{0}) - B(p_{0})$ contain the
contributions  from the positive-parity   and negative-parity triply heavy baryon states,  respectively.

 In the following, we  briefly outline the operator product expansion performed  at the large space-like region $p^2 \ll 0$.
 We contract the heavy quark fields in the correlation functions $\Pi^{\pm}(p)$ and $\Pi^{\pm}_{\mu\nu}(p)$ with
Wick theorem,  substitute the full  heavy quark propagators into the
correlation functions  $\Pi^{\pm}(p)$ and $\Pi^{\pm}_{\mu\nu}(p)$ firstly, then  complete  the integrals in the
coordinate space and momentum
space sequentially to obtain the correlation functions $\Pi^{\pm}(p)$ and $\Pi^{\pm}_{\mu\nu}(p)$ at the quark level.
In calculations, we take into account all diagrams like the typical
ones shown in Fig.1. Once the analytical  quark-level correlation functions  $\Pi^{\pm}(p)$ and $\Pi^{\pm}_{\mu\nu}(p)$ are obtained,
we  take the limit $\vec{p} = 0$,  and
use the dispersion relation to obtain the QCD spectral densities
$\rho^A(p_0)$ and $\rho^B(p_0)$ (which correspond to the tensor
structures $\gamma_0$ and $1$ respectively). Finally we introduce the weight
functions $\exp\left[-\frac{p_0^2}{T^2}\right]$,
$p_0^2\exp\left[-\frac{p_0^2}{T^2}\right]$,   and obtain the
following QCD sum rules,
\begin{eqnarray}
  \lambda_{\pm}^2\exp\left[-\frac{M_{\pm}^2}{T^2}\right]&=&\int_{\Delta}^{\sqrt{s_0}}dp_0\left[\rho^A(p_0)\pm\rho^B(p_0)\right]\exp\left[-\frac{p_0^2}{T^2}\right] \, ,
\end{eqnarray}
\begin{eqnarray}
 \lambda_{\pm}^2M_{\pm}^2\exp\left[-\frac{M_{\pm}^2}{T^2}\right]&=&\int_{\Delta}^{\sqrt{s_0}}dp_0\left[\rho^A(p_0)\pm\rho^B(p_0)\right]p_0^2\exp\left[-\frac{p_0^2}{T^2}\right]\,,
\end{eqnarray}
where  the $s_0$ are the continuum threshold parameters, the $T^2$ are the Borel
parameters, and the threshold parameters  $\Delta=2m_Q+m_{Q'}$ or  $3m_Q$. The spectral
densities $\rho^A(p_0)$ and $\rho^B(p_0)$ at the level of
quark-gluon degrees of freedom are given explicitly in the Appendix. We can obtain the  masses $M_{\pm}$  and pole residues
$\lambda_{\pm}$ by solving above equations with simultaneous  iterations.

\begin{figure}
 \centering
 \includegraphics[totalheight=3cm,width=14cm]{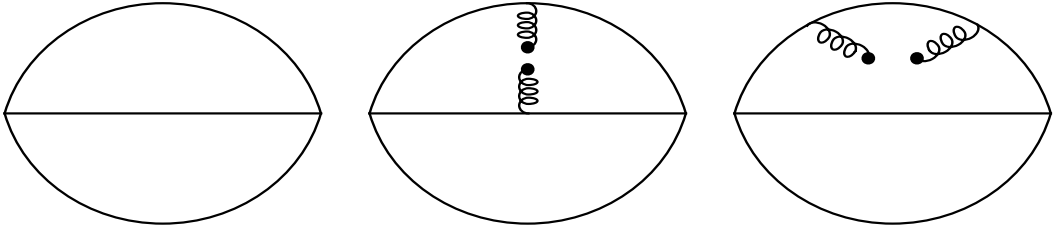}
    \caption{The typical diagrams we calculate in the operator product expansion, we take into account
    the tree-level perturbative term and the gluon condensates. }
\end{figure}

\section{Numerical results and discussions}
The input parameters are taken as $\langle \frac{\alpha_s GG}{\pi}\rangle=(0.012 \pm
0.004)\,\rm{GeV}^4 $ \cite{LCSRreview},
 $m_c=1.3\,\rm{GeV}$ and $m_b=4.7\,\rm{GeV}$ \cite{PDG}.
The value of the gluon condensate $\langle \frac{\alpha_s
GG}{\pi}\rangle $ has been updated from time to time, and changes
greatly \cite{NarisonBook}.
 The updated value $\langle \frac{\alpha_s GG}{\pi}\rangle=(0.023 \pm
0.003)\,\rm{GeV}^4 $ \cite{NarisonBook} and the standard value
$\langle \frac{\alpha_s GG}{\pi}\rangle=(0.012 \pm
0.004)\,\rm{GeV}^4 $ \cite{LCSRreview} lead to a tiny  difference
 as  the gluon condensate  makes tiny  contribution.
The  heavy quark masses appearing in the perturbative terms  are
usually taken to be the pole masses in the QCD sum rules, while the
choice of the $m_Q$ in the leading-order coefficients of the
higher-dimensional terms is arbitrary \cite{NarisonBook,Kho9801}. In calculations, we observe that
the dominating contributions come from the perturbative term. So we take the pole masses and neglect
 the uncertainties of the pole masses. The integral intervals of the energy $p_0$ are rather small,
variations of the threshold parameters $\Delta=(2m_Q+m_{Q'})$ or $3m_Q$ can lead to remarkable
changes of the continuum threshold parameters $\sqrt{s_0}$,  we can fix the $\Delta$ and vary the $\sqrt{s_0}$.

In the conventional QCD sum rules \cite{SVZ79}, there are two
criteria (pole dominance and convergence of the operator product
expansion) for choosing  the Borel parameter $T^2$ and continuum threshold
parameter $s_0$.  We impose the two criteria on the triply
heavy baryon states to choose the Borel parameter $T^2$ and continuum
threshold parameter $s_0$. In our previous works on the heavy and doubly heavy baryon states, the pole contributions are taken  as
$(45-80)\%$  \cite{H-baryon-Wang}. We can take the same pole contributions, then search for the continuum threshold parameters
 $\sqrt{s_0}$ to reproduce the relation $\sqrt{s_0}=M_{\pm}+(0.4\sim0.6)\,\rm{GeV}$. The Borel parameters $T^2$,
 continuum threshold parameters $\sqrt{s_0}$, masses, pole residues are  shown in Table 1 and Figs.2-3.
In this article, we have neglected the contributions of the perturbative $\mathcal {O}(\alpha_s)$  corrections, which can be taken into account by introducing  formal coefficient $1+\frac{\alpha_s}{\pi}f(m_Q,m_{Q'},s_0)$ through the unknown function $f(m_Q,m_{Q'},s_0)$. As the  dominant contributions come from the perturbative term, we expect that the  $\mathcal {O}(\alpha_s)$  corrections to the perturbative term cannot change the masses remarkably,  those effects  can be absorbed in the pole residues approximately.

If we choose  the structures $\gamma_0$ and $1$  to study the masses,
there are contaminations of the negative- (or positive-) parity triply heavy
baryon states to the positive- (or negative-) parity triply heavy baryon states,   the corresponding  fractions can be expressed as
\begin{eqnarray}
 R_{\pm}&=& \frac{\int_{\Delta}^{\sqrt{s_0}}dp_0\left[\rho^A(p_0) \mp\rho^B(p_0)\right]\exp\left[-\frac{p_0^2}{T^2}\right]}
{\int_{\Delta}^{\sqrt{s_0}}dp_0\left[\rho^A(p_0)
\pm\rho^B(p_0)\right]\exp\left[-\frac{p_0^2}{T^2}\right]} \, .
\end{eqnarray}
In this article, we separate the contributions of the positive-parity and negative-parity triply heavy baryon states explicitly.

In calculations, we have taken the pole masses. On the other hand, we can take the   $\overline{MS}$ masses $m_c(m_c^2)=1.2\,\rm{GeV}$, $m_b(m_b^2)=4.2\,\rm{GeV}$,  as the $\overline{MS}$ masses are also used in the QCD sum rules,  for example, in studying  the $B\to \pi $ form-factors \cite{Bpi}. We choose the same  Borel parameters and suitable continuum threshold parameters to reproduce the same pole contributions so as to obtain  the ground state masses and pole residues, the predictions  are  presented in Table 1. From Table 1, we can see that the pole masses and the $\overline{MS}$ masses result in large discrepancies   for the masses of the   triply heavy baryon states. If the $\overline{MS}$ masses are taken,  the present predictions are compatible with the values from Ref.\cite{ZhanM}   within uncertainties.  In Ref.\cite{ZhanM}, the contributions of the positive-parity baryon states are   not distinguished from that of the negative-parity baryon states.
 For the established bottom baryon states $\Sigma_b$ with three stars, the masses are $M_{\Sigma_b^+}=5.8078\,\rm{GeV}$ and $M_{\Sigma_b^-}=5.8152\,\rm{GeV}$ respectively from the Particle Data Group \cite{PDG}. The prediction $M_{\Sigma_b}=(5.80\pm0.19)\,\rm{GeV}$ based on the pole mass $m_b =(4.7\pm0.1)\,\rm{GeV}$ is consistent with experimental data \cite{H-baryon-Wang}, while the prediction $M_{\Sigma_b}=(5.72\pm0.19)\,\rm{GeV}$ based on the $\overline{MS}$ mass $m_b(m_b^2)=(4.2\pm0.1)\,\rm{GeV}$  underestimates
  the experimental data about $80\,\rm{MeV}$,  if we take the same values of other parameters; so the pole masses are preferred. In the QCD sum rules, if the variations of the threshold parameters $\Delta$ can   lead to relatively large changes for the integral ranges  $\sqrt{s_0}-\Delta$, the predictions are  sensitive  to the masses $m_Q$. In the present case,   the mass uncertainty $\delta m_b=0.1\,\rm{GeV}$ can result in uncertainty $\frac{\delta\Delta}{\sqrt{s_0}-\Delta}\approx 20\%$ for the triply-bottom baryon state $\Omega_{bbb}$. The pole mass and the $\overline{MS}$ mass correspond to quite different continuum threshold parameters $s_0$, see Table 1.
  On the other hand, if the variations of the threshold parameters $\Delta$  result in small values of the $\frac{\delta\Delta}{\sqrt{s_0}-\Delta}$, the predictions based on the pole masses and the $\overline{MS}$ masses  lead to small discrepancies. Irrespective of the pole masses and the $\overline{MS}$ masses,  it would be better to understand the heavy quark masses $m_Q$  as the effective
quark masses (or just the mass parameters). Our previous works on
the mass spectrum of the ${\frac{1}{2}}^\pm$ and ${\frac{3}{2}}^\pm$
heavy and doubly heavy baryon states  indicate such parameters (the pole masses) can lead to
satisfactory results \cite{H-baryon-Wang},  we prefer   the pole masses.

There are no experimental data for the masses of the triply heavy baryon states, the present predictions
are compared with other theoretical calculations, such as     the  QCD bag model \cite{HaseM}, the quark model estimation
\cite{BjorM},  the variational approach \cite{JiaM}, the modified bag model
\cite{BeroM}, the relativistic three-quark model \cite{MartM}, the QCD sum rules \cite{ZhanM},
the non-relativistic three-quark model \cite{RobeM}, see Table 2.
All those predictions should be confronted with the experimental data in the future.
The LHC will be the world's most copious  source of the $b$  hadrons,
and  a complete spectrum of the $b$ and $c$ hadrons will be available
through the gluon fusions. In proton-proton collisions at
$\sqrt{s}=14\,\rm{TeV}$, the $b\bar{b}$ cross section is expected to
be $\sim 500\mu b$ producing $10^{12}$ $b\bar{b}$ pairs in a
standard  year of running at the LHCb operational luminosity of
$2\times10^{32} \rm{cm}^{-2} \rm{sec}^{-1}$ \cite{LHC}.

\begin{table}
\begin{center}
\begin{tabular}{|c|c|c|c|c|c|c|c|}\hline\hline
                                               & $T^2 (\rm{GeV}^2)$  & $\sqrt{s_0} (\rm{GeV})$ & pole         & $M(\rm{GeV})$    & $\lambda(\rm{GeV}^3)$   \\ \hline
             $\Omega_{ccc}({\frac{3}{2}^+})$   & $4.6-6.4$           & $5.6\pm0.2$             & $(41-79)\%$  & $4.99\pm0.14$    & $0.20\pm0.04$          \\ \hline
             $\Omega_{ccb}({\frac{1}{2}^+})$   & $6.3-8.3$           & $8.8\pm0.2$             & $(43-80)\%$  & $8.23\pm0.13$    & $0.47\pm0.10$        \\ \hline
             $\Omega_{ccb}({\frac{3}{2}^+})$   & $6.4-8.4$           & $8.8\pm0.2$             & $(43-80)\%$  & $8.23\pm0.13$    & $0.26\pm0.05$        \\ \hline
             $\Omega_{bbc}({\frac{1}{2}^+})$   & $8.0-10.0$          & $12.0\pm0.2$            & $(44-79)\%$  & $11.50\pm0.11$   & $0.68\pm0.15$        \\ \hline
             $\Omega_{bbc}({\frac{3}{2}^+})$   & $8.0-10.0$          & $12.0\pm0.2$            & $(45-80)\%$  & $11.49\pm0.11$   & $0.39\pm0.09$        \\ \hline            $\Omega_{bbb}({\frac{3}{2}^+})$   & $10.0-12.0$         & $15.3\pm0.2$            & $(45-79)\%$  & $14.83\pm0.10$   & $0.68\pm0.16$        \\ \hline
             $\Omega_{ccc}({\frac{3}{2}^-})$   & $5.1-7.1$           & $5.8\pm0.2$             & $(44-80)\%$  & $5.11\pm0.15$    & $0.24\pm0.04$        \\ \hline
             $\Omega_{ccb}({\frac{1}{2}^-})$   & $7.2-9.2$           & $9.0\pm0.2$             & $(46-79)\%$  & $8.36\pm0.13$    & $0.57\pm0.11$        \\ \hline
             $\Omega_{ccb}({\frac{3}{2}^-})$   & $7.3-9.3$           & $9.0\pm0.2$             & $(47-79)\%$  & $8.36\pm0.13$    & $0.32\pm0.06$        \\ \hline
             $\Omega_{bbc}({\frac{1}{2}^-})$   & $9.5-11.5$          & $12.2\pm0.2$            & $(46-77)\%$  & $11.62\pm0.11$   & $0.86\pm0.17$        \\ \hline
             $\Omega_{bbc}({\frac{3}{2}^-})$   & $9.5-11.5$          & $12.2\pm0.2$            & $(47-78)\%$  & $11.62\pm0.11$   & $0.49\pm0.10$        \\ \hline
             $\Omega_{bbb}({\frac{3}{2}^-})$   & $11.4-14.0$         & $15.5\pm0.2$            & $(48-80)\%$  & $14.95\pm0.11$   & $0.86\pm0.17$        \\ \hline
  $\overline{\Omega}_{ccc}({\frac{3}{2}^+})$   & $4.6-6.4$           & $5.4\pm0.2$             & $(42-79)\%$  & $4.76\pm0.14$    & $0.20\pm0.04$          \\ \hline
  $\overline{\Omega}_{ccb}({\frac{1}{2}^+})$   & $6.3-8.3$           & $8.2\pm0.2$             & $(42-78)\%$  & $7.61\pm0.13$    & $0.47\pm0.10$        \\ \hline
  $\overline{\Omega}_{ccb}({\frac{3}{2}^+})$   & $6.4-8.4$           & $8.2\pm0.2$             & $(43-79)\%$  & $7.60\pm0.13$    & $0.26\pm0.05$        \\ \hline
  $\overline{\Omega}_{bbc}({\frac{1}{2}^+})$   & $8.0-10.0$          & $11.0\pm0.2$            & $(43-78)\%$  & $10.47\pm0.12$   & $0.68\pm0.15$        \\ \hline
  $\overline{\Omega}_{bbc}({\frac{3}{2}^+})$   & $8.0-10.0$          & $11.0\pm0.2$            & $(44-78)\%$  & $10.46\pm0.12$   & $0.39\pm0.09$        \\ \hline                     $\overline{\Omega}_{bbb}({\frac{3}{2}^+})$   & $10.0-12.0$         & $13.9\pm0.2$            & $(45-78)\%$  & $13.40\pm0.10$   & $0.66\pm0.15$        \\ \hline
  $\overline{\Omega}_{ccc}({\frac{3}{2}^-})$   & $5.1-7.1$           & $5.6\pm0.2$             & $(44-80)\%$  & $4.88\pm0.15$    & $0.24\pm0.04$        \\ \hline
  $\overline{\Omega}_{ccb}({\frac{1}{2}^-})$   & $7.2-9.2$           & $8.4\pm0.2$             & $(45-78)\%$  & $7.74\pm0.13$    & $0.57\pm0.11$        \\ \hline
  $\overline{\Omega}_{ccb}({\frac{3}{2}^-})$   & $7.3-9.3$           & $8.4\pm0.2$             & $(46-78)\%$  & $7.73\pm0.13$    & $0.32\pm0.06$        \\ \hline
  $\overline{\Omega}_{bbc}({\frac{1}{2}^-})$   & $9.5-11.5$          & $11.2\pm0.2$            & $(45-75)\%$  & $10.60\pm0.12$   & $0.84\pm0.17$        \\ \hline
  $\overline{\Omega}_{bbc}({\frac{3}{2}^-})$   & $9.5-11.5$          & $11.2\pm0.2$            & $(46-76)\%$  & $10.59\pm0.11$   & $0.47\pm0.10$        \\ \hline          $\overline{\Omega}_{bbb}({\frac{3}{2}^-})$   & $11.4-14.0$         & $14.1\pm0.2$            & $(46-78)\%$  & $13.52\pm0.11$   & $0.82\pm0.16$        \\ \hline
 \hline
\end{tabular}
\end{center}
\caption{ The Borel parameters, continuum threshold parameters, pole contributions, masses and pole residues of  the triply
heavy baryon states. The overline on the $\Omega_{QQQ'}$ denotes the $\overline{MS}$ masses are used. }
\end{table}

\begin{figure}
 \centering
 \includegraphics[totalheight=4.3cm,width=4.7cm]{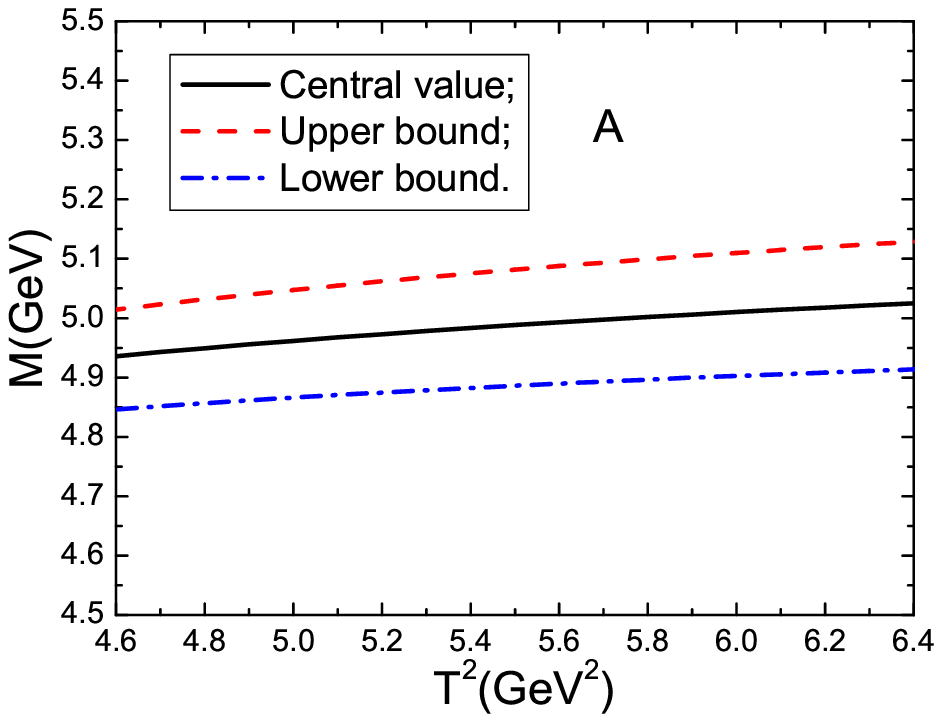}
 \includegraphics[totalheight=4.3cm,width=4.7cm]{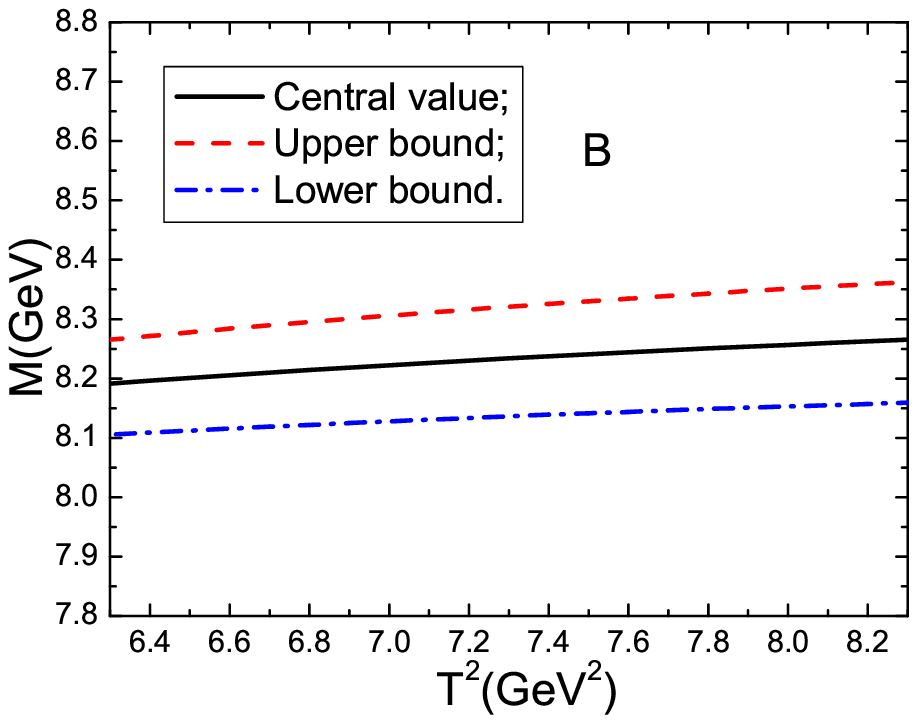}
 \includegraphics[totalheight=4.3cm,width=4.7cm]{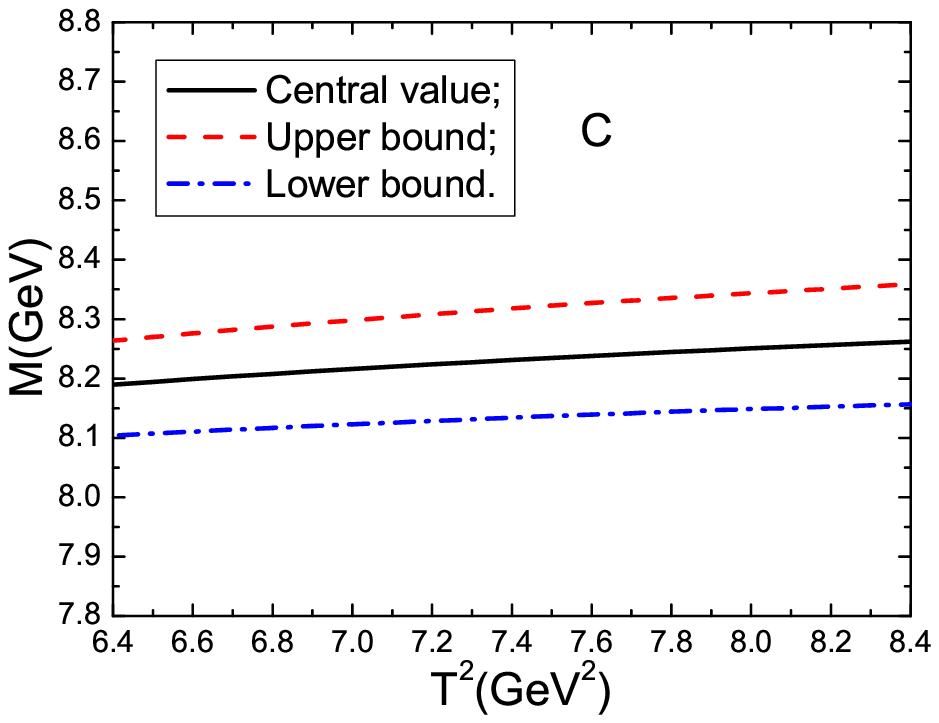}
  \includegraphics[totalheight=4.3cm,width=4.7cm]{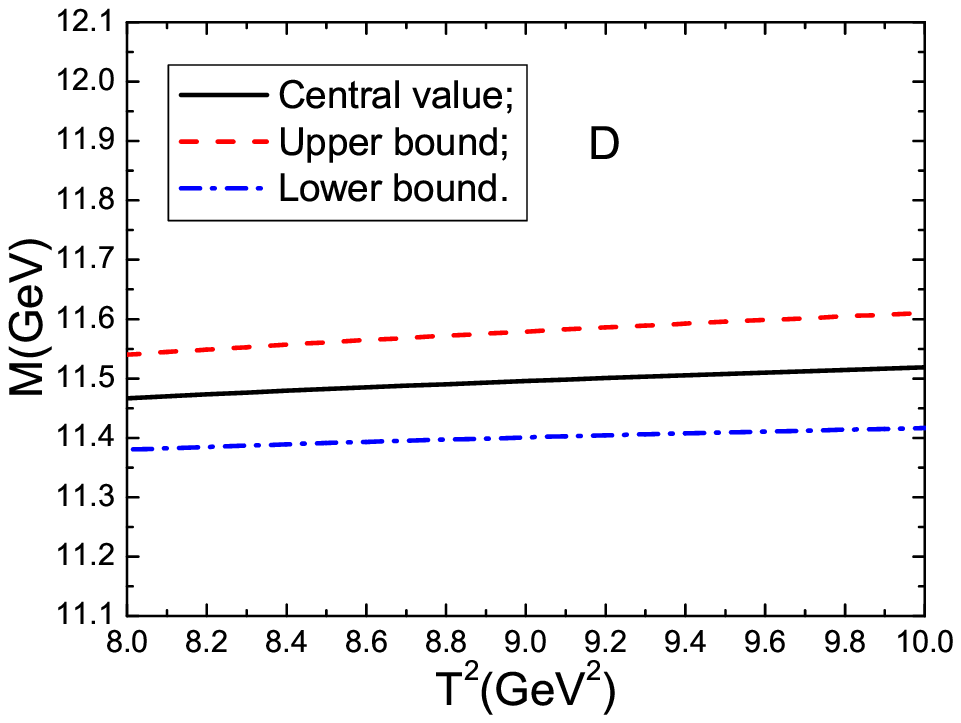}
 \includegraphics[totalheight=4.3cm,width=4.7cm]{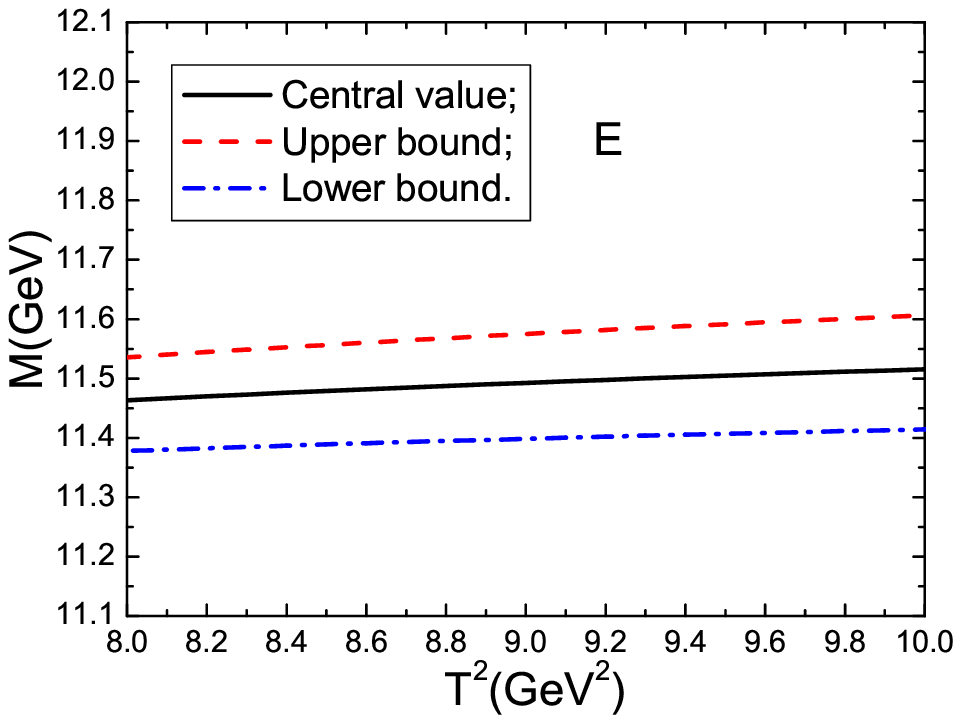}
 \includegraphics[totalheight=4.3cm,width=4.7cm]{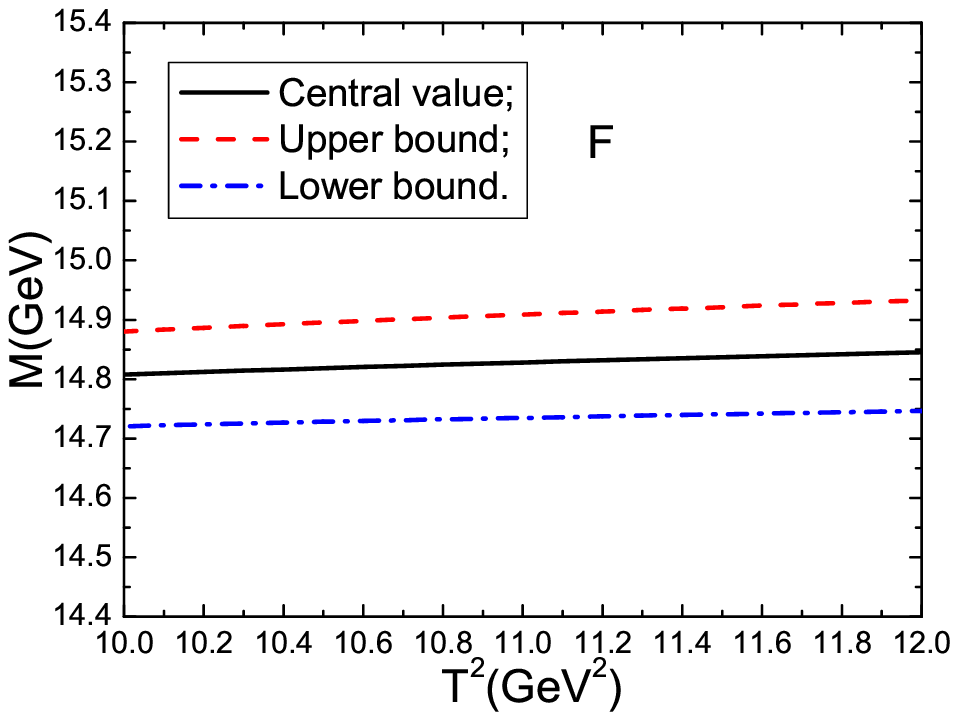}
  \includegraphics[totalheight=4.3cm,width=4.7cm]{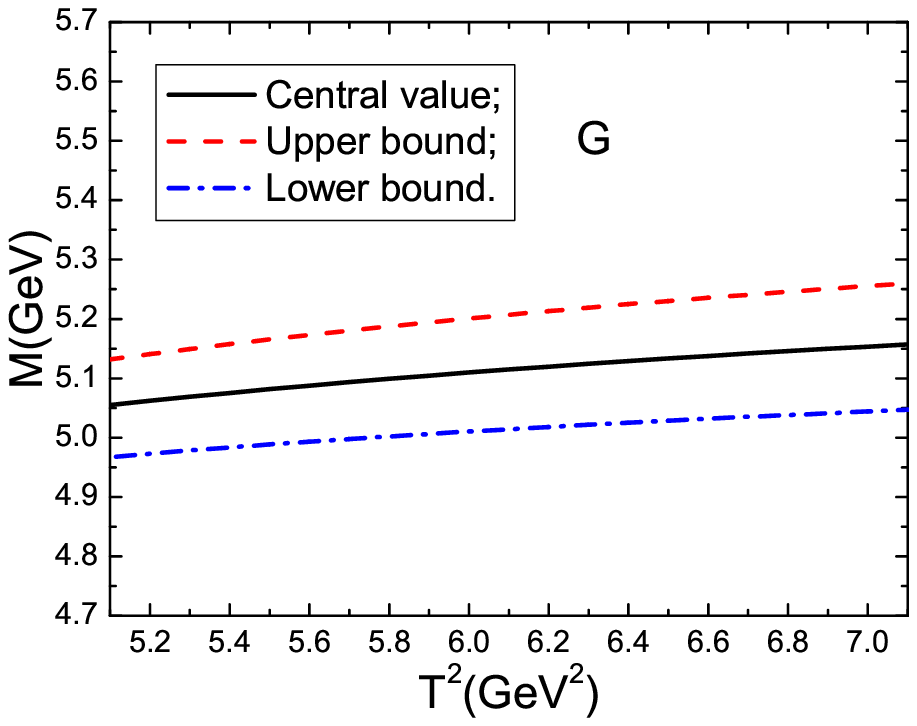}
 \includegraphics[totalheight=4.3cm,width=4.7cm]{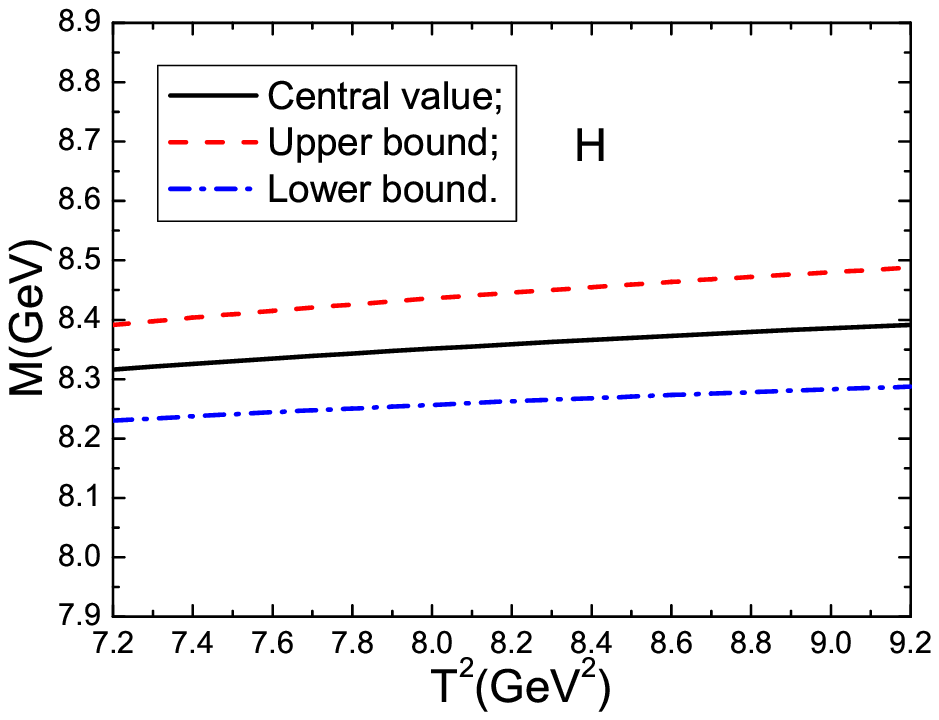}
 \includegraphics[totalheight=4.3cm,width=4.7cm]{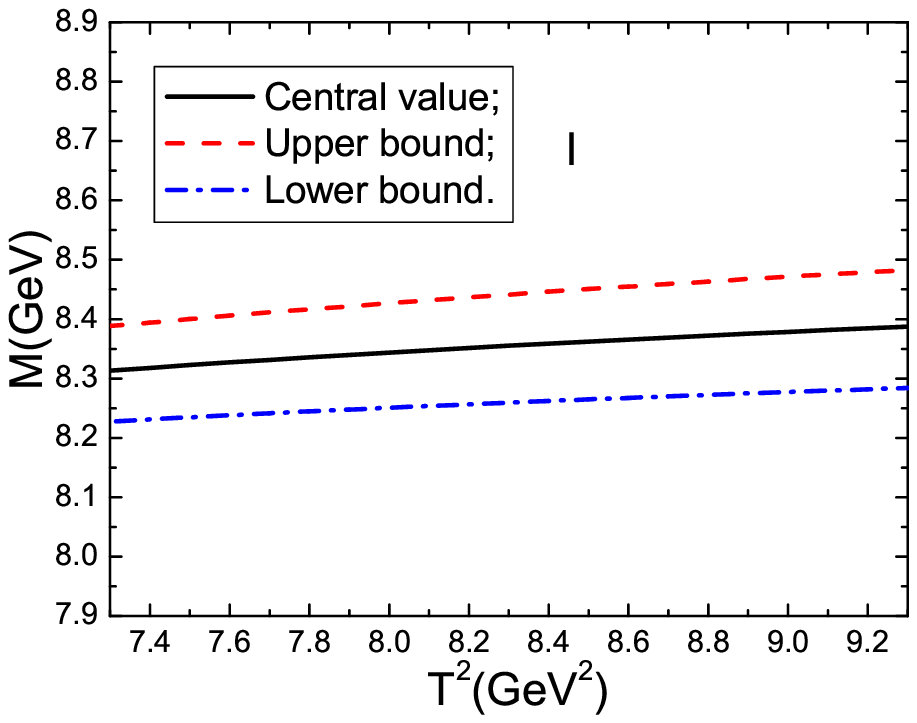}
  \includegraphics[totalheight=4.3cm,width=4.7cm]{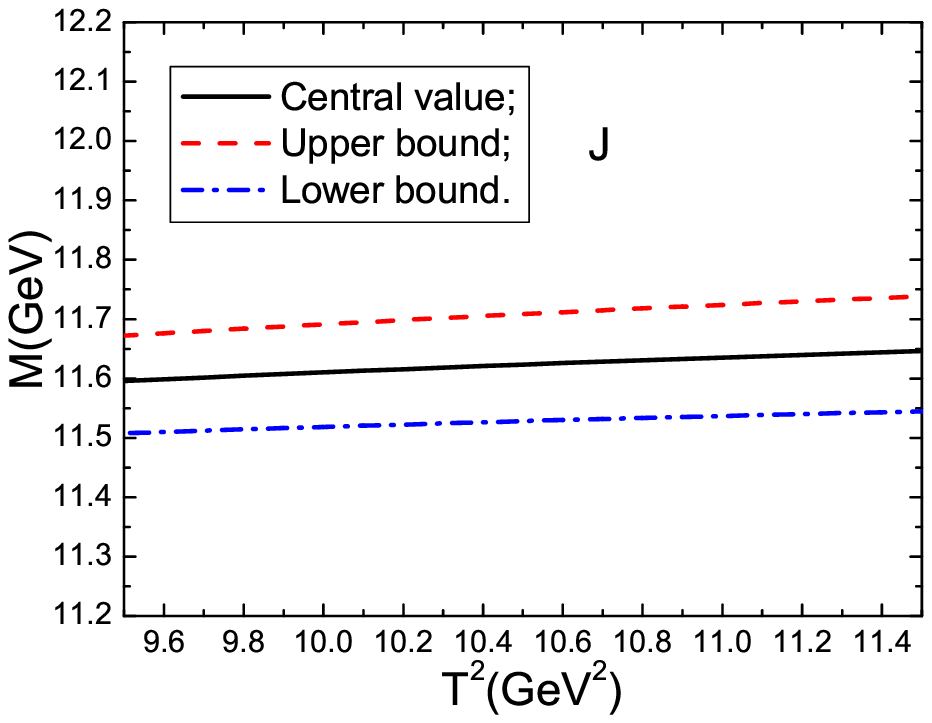}
 \includegraphics[totalheight=4.3cm,width=4.7cm]{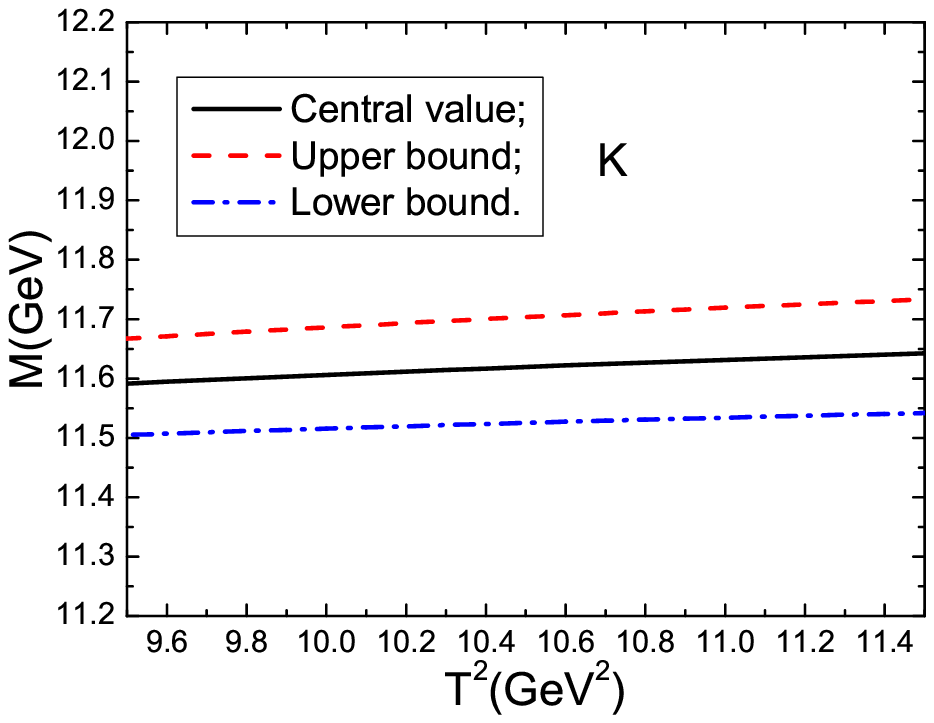}
 \includegraphics[totalheight=4.3cm,width=4.7cm]{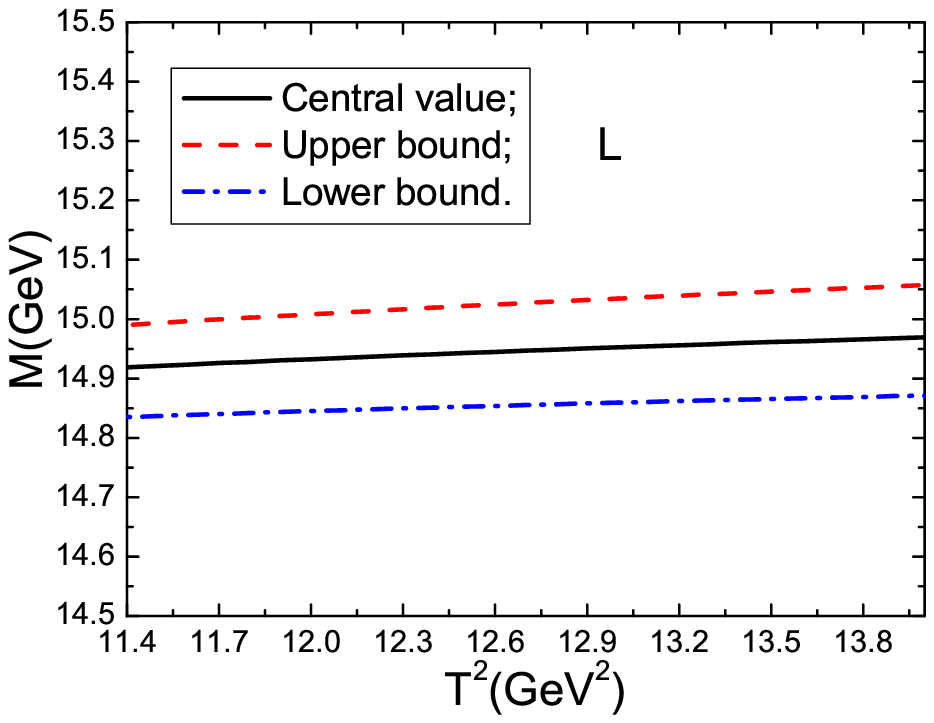}
   \caption{ The  masses of the   triply  heavy baryon states with variations of the Borel parameters,
   the $A$, $B$, $C$, $D$, $E$, $F$, $G$, $H$, $I$, $J$, $K$ and $L$  correspond
     to the  $\Omega_{ccc}({3\over 2}^+)$, $\Omega_{ccb}({1\over 2}^+)$, $\Omega_{ccb}({3\over 2}^+)$, $\Omega_{bbc}({1\over 2}^+)$, $\Omega_{bbc}({3\over 2}^+)$, $\Omega_{bbb}({3\over 2}^+)$,  $\Omega_{ccc}({3\over 2}^-)$, $\Omega_{ccb}({1\over 2}^-)$, $\Omega_{ccb}({3\over 2}^-)$, $\Omega_{bbc}({1\over 2}^-)$, $\Omega_{bbc}({3\over 2}^-)$ and $\Omega_{bbb}({3\over 2}^-)$, respectively.  }
\end{figure}

\begin{figure}
 \centering
 \includegraphics[totalheight=4.3cm,width=4.7cm]{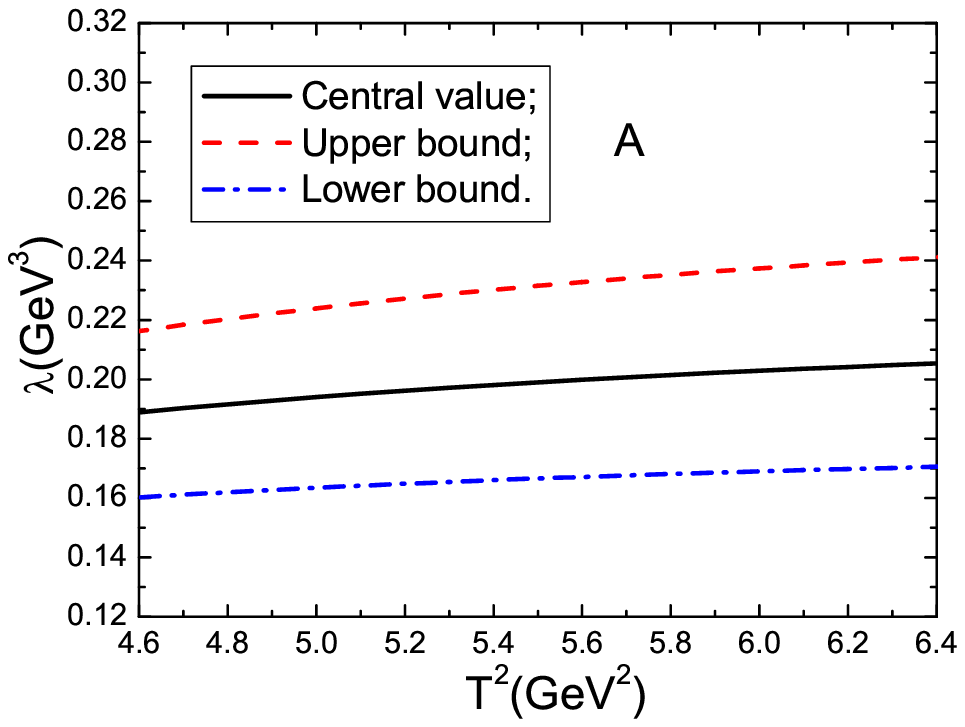}
 \includegraphics[totalheight=4.3cm,width=4.7cm]{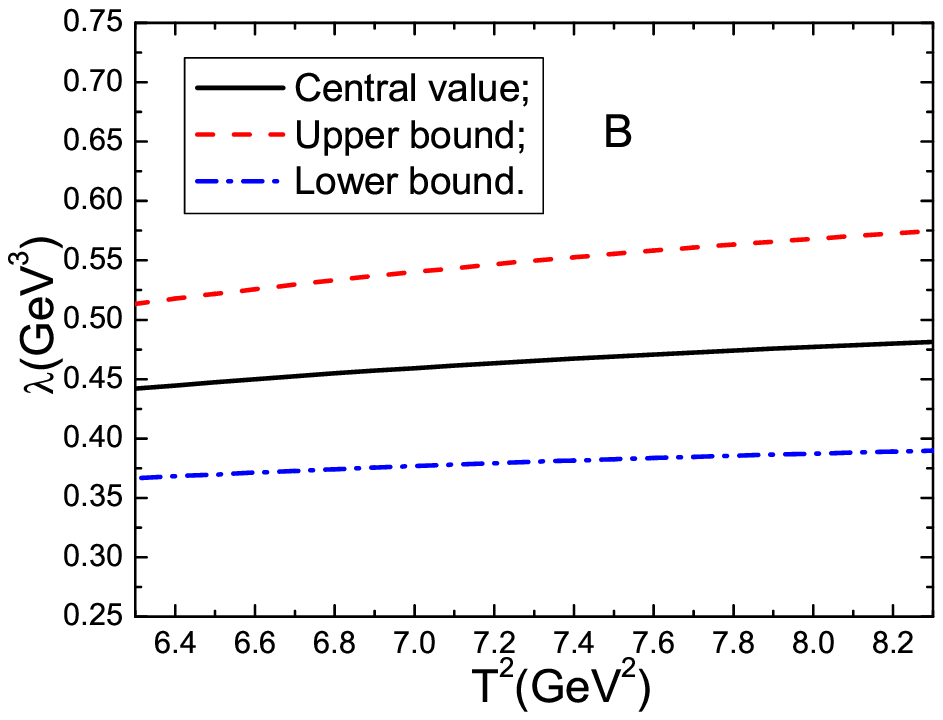}
 \includegraphics[totalheight=4.3cm,width=4.7cm]{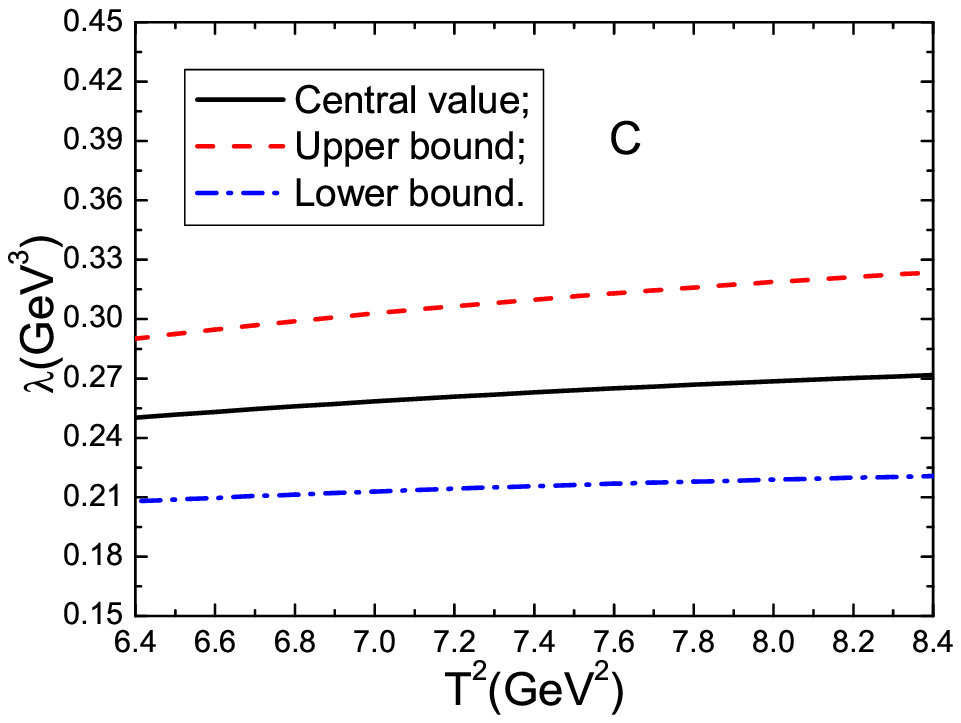}
 \includegraphics[totalheight=4.3cm,width=4.7cm]{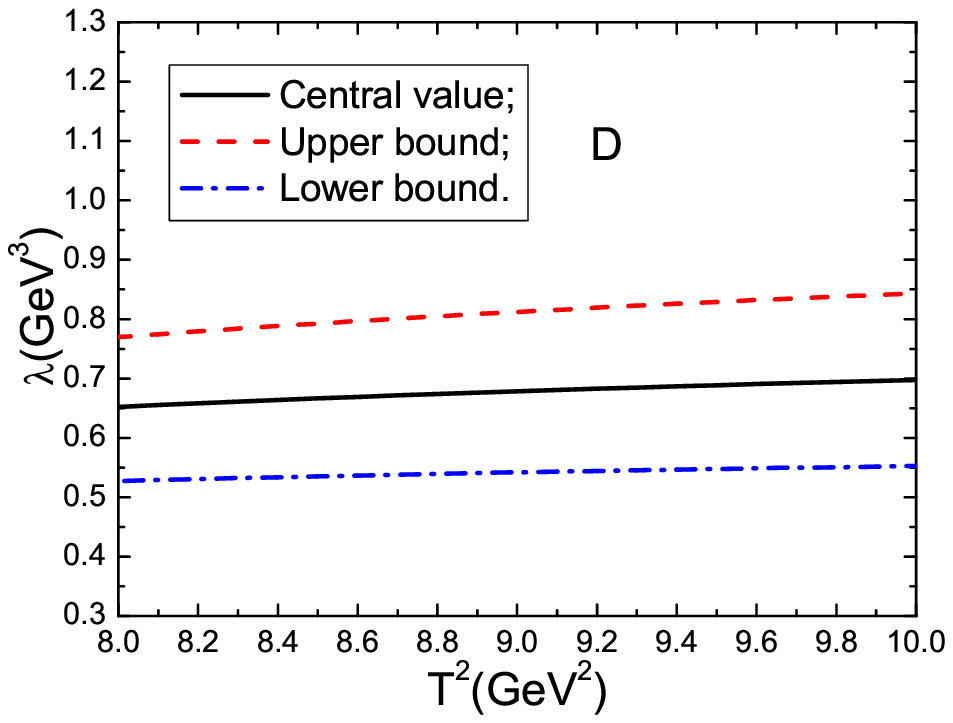}
 \includegraphics[totalheight=4.3cm,width=4.7cm]{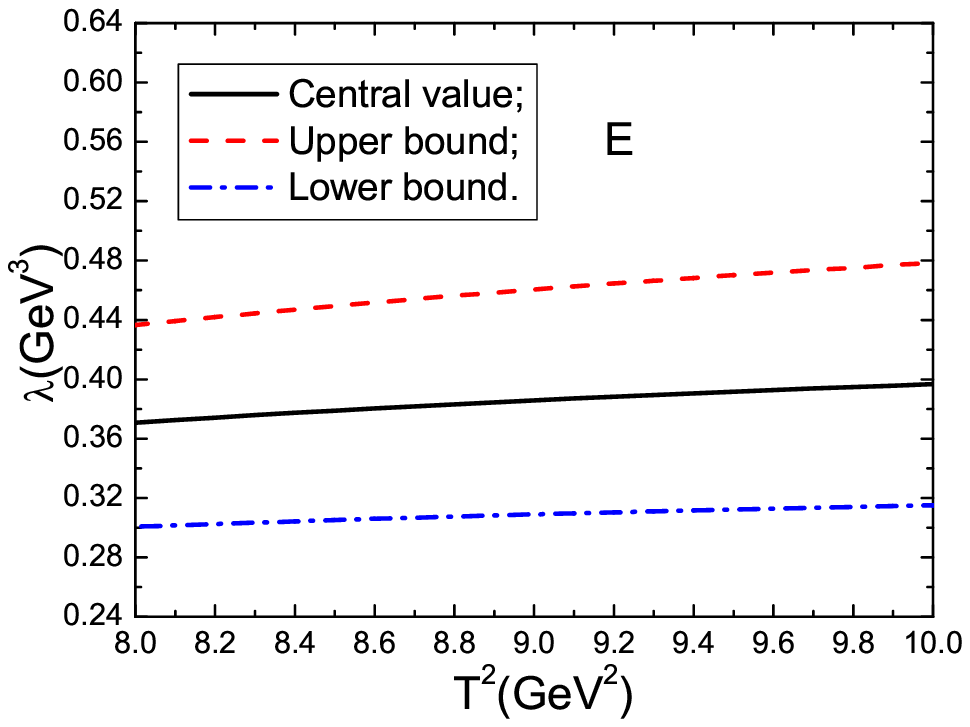}
 \includegraphics[totalheight=4.3cm,width=4.7cm]{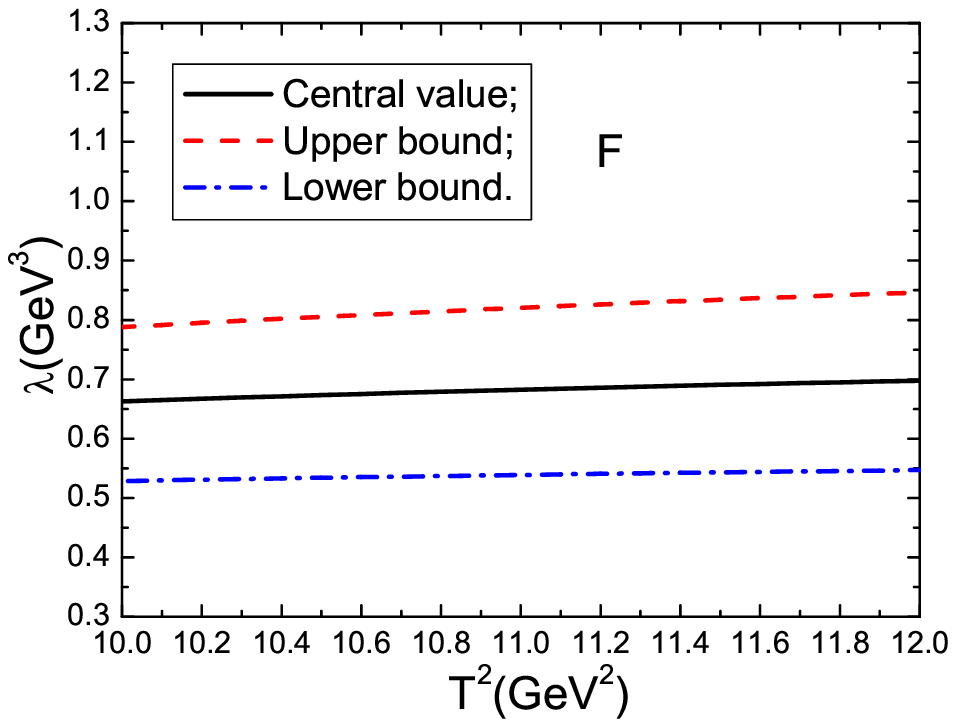}
 \includegraphics[totalheight=4.3cm,width=4.7cm]{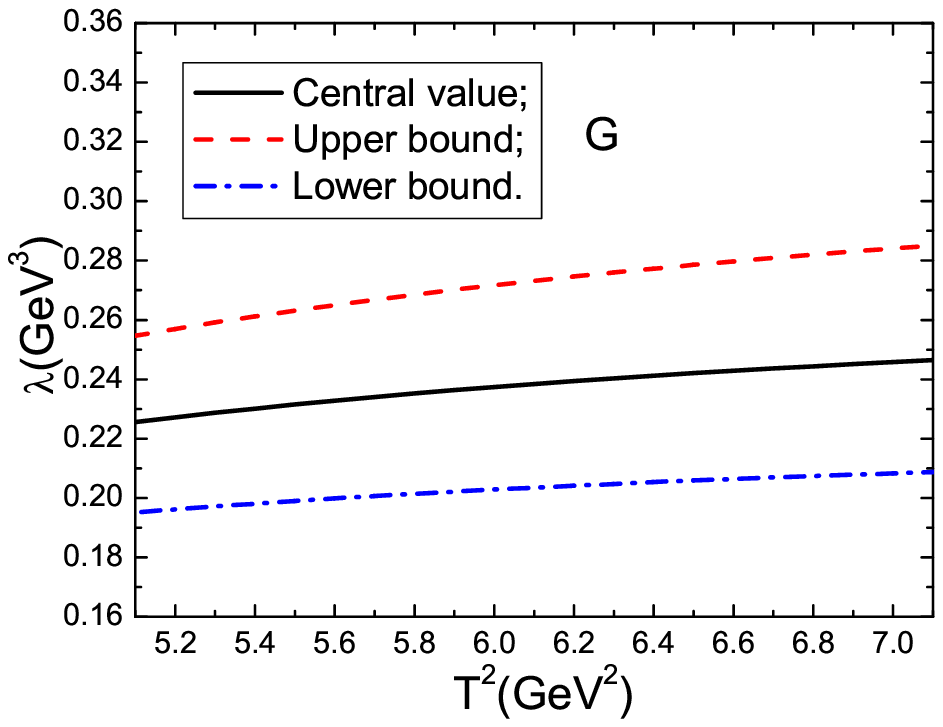}
 \includegraphics[totalheight=4.3cm,width=4.7cm]{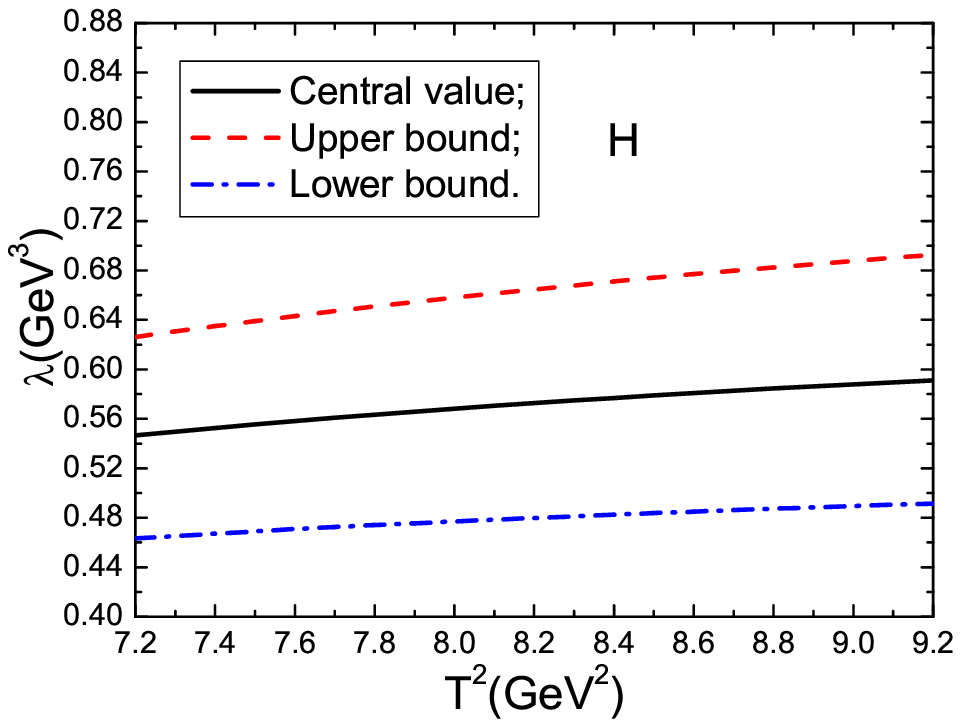}
 \includegraphics[totalheight=4.3cm,width=4.7cm]{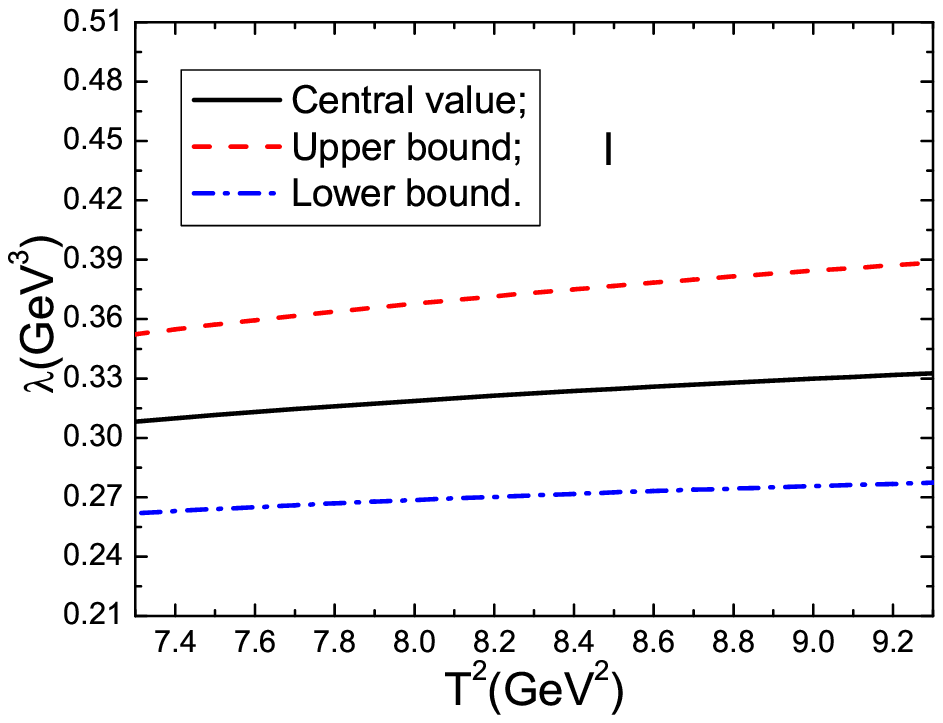}
 \includegraphics[totalheight=4.3cm,width=4.7cm]{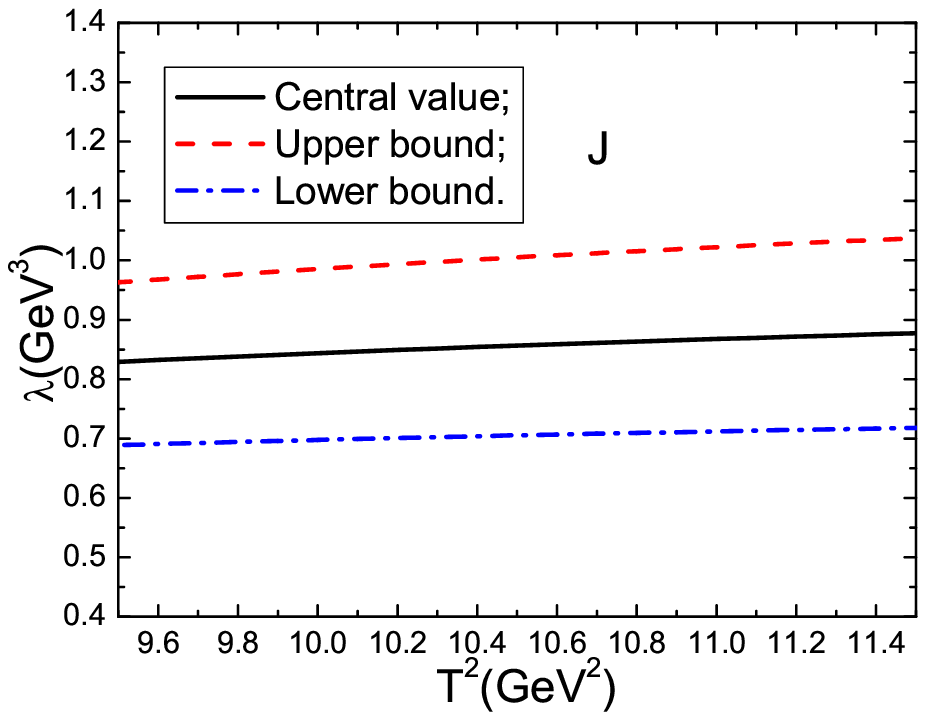}
 \includegraphics[totalheight=4.3cm,width=4.7cm]{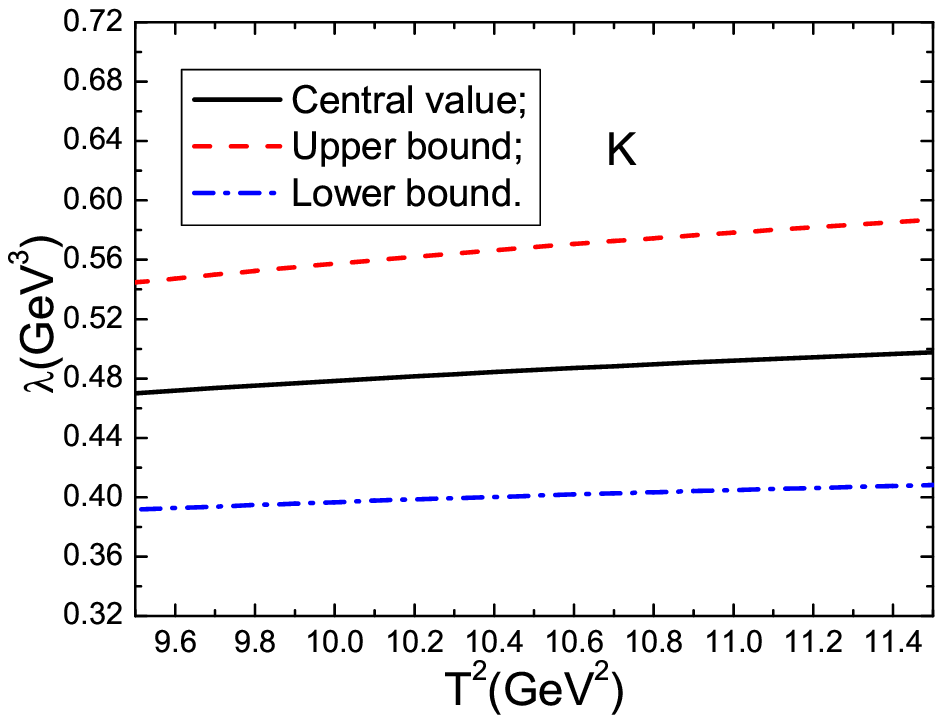}
 \includegraphics[totalheight=4.3cm,width=4.7cm]{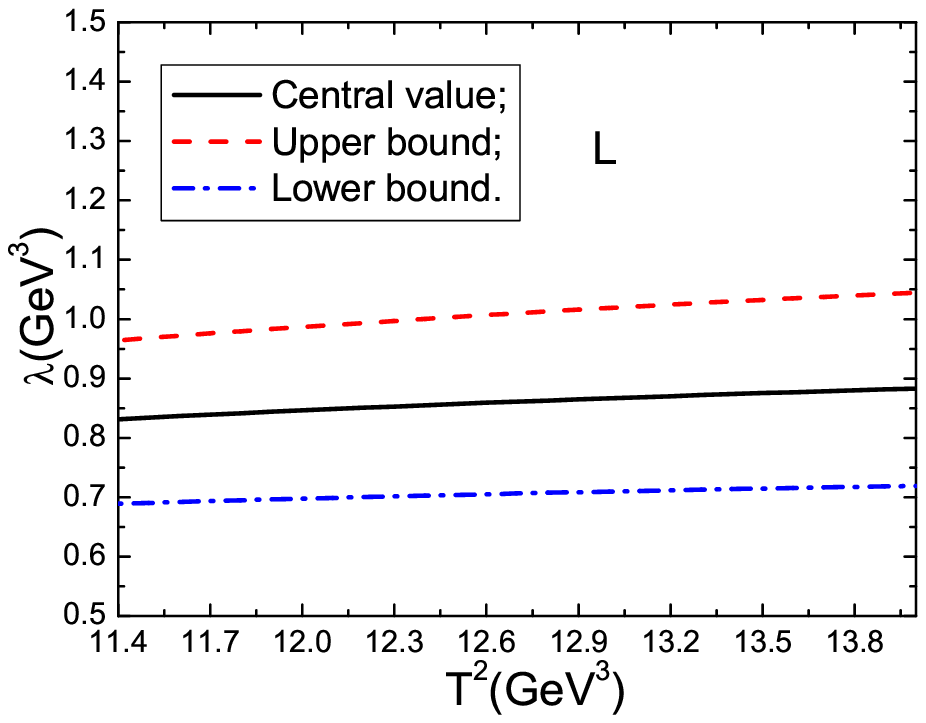}
  \caption{ The  pole residues  of the   triply  heavy baryon states with variations of the Borel parameters,
  the $A$, $B$, $C$, $D$, $E$, $F$, $G$, $H$, $I$, $J$, $K$ and $L$  correspond
     to  the $\Omega_{ccc}({3\over 2}^+)$, $\Omega_{ccb}({1\over 2}^+)$, $\Omega_{ccb}({3\over 2}^+)$, $\Omega_{bbc}({1\over 2}^+)$, $\Omega_{bbc}({3\over 2}^+)$, $\Omega_{bbb}({3\over 2}^+)$,  $\Omega_{ccc}({3\over 2}^-)$, $\Omega_{ccb}({1\over 2}^-)$, $\Omega_{ccb}({3\over 2}^-)$, $\Omega_{bbc}({1\over 2}^-)$, $\Omega_{bbc}({3\over 2}^-)$ and $\Omega_{bbb}({3\over 2}^-)$, respectively.   }
\end{figure}

\begin{table}
\begin{center}
\begin{tabular}{|c|c|c|c|c|c|c|c|c|}\hline\hline
                                    & This work      &\cite{HaseM} &\cite{BjorM} & \cite{JiaM}  &\cite{BeroM} &\cite{MartM} &\cite{ZhanM}     &\cite{RobeM}\\ \hline
   $\Omega_{ccc}({\frac{3}{2}^+})$  &$4.99\pm0.14$   & 4.79        & 4.925       & 4.76         & 4.777       & 4.803       & $4.67\pm0.15$   & 4.965 \\ \hline
   $\Omega_{ccb}({\frac{1}{2}^+})$  &$8.23\pm0.13$   &             &             &              & 7.984       & 8.018       & $7.41\pm0.13$   & 8.245  \\ \hline
   $\Omega_{ccb}({\frac{3}{2}^+})$  &$8.23\pm0.13$   & 8.03        & 8.200       & 7.98         & 8.005       & 8.025       & $7.45\pm0.16$   & 8.265 \\ \hline
   $\Omega_{bbc}({\frac{1}{2}^+})$  &$11.50\pm0.11$  &             &             &              & 11.139      & 11.280      & $10.30\pm0.10$  & 11.535 \\ \hline
   $\Omega_{bbc}({\frac{3}{2}^+})$  &$11.49\pm0.11$  & 11.20       & 11.480      & 11.19        & 11.163      & 11.287      & $10.54\pm0.11$  & 11.554 \\ \hline
   $\Omega_{bbb}({\frac{3}{2}^+})$  &$14.83\pm0.10$  & 14.30       & 14.760      & 14.37        & 14.276      & 14.569      & $13.28\pm0.10$  & 14.834 \\ \hline
   $\Omega_{ccc}({\frac{3}{2}^-})$  &$5.11\pm0.15$   &             &             &              &             &             &                 & 5.160 \\ \hline
   $\Omega_{ccb}({\frac{1}{2}^-})$  &$8.36\pm0.13$   &             &             &              &             &             &                 & 8.418 \\ \hline
   $\Omega_{ccb}({\frac{3}{2}^-})$  &$8.36\pm0.13$   &             &             &              &             &             &                 & 8.420 \\ \hline
   $\Omega_{bbc}({\frac{1}{2}^-})$  &$11.62\pm0.11$  &             &             &              &             &             &                 & 11.710 \\ \hline
   $\Omega_{bbc}({\frac{3}{2}^-})$  &$11.62\pm0.11$  &             &             &              &             &             &                 & 11.711 \\ \hline
   $\Omega_{bbb}({\frac{3}{2}^-})$  &$14.95\pm0.11$  &             &             &              &             &             &                 & 14.976\\ \hline
                      \hline
\end{tabular}
\end{center}
\caption{ The masses of the triply heavy baryon states compared with other theoretical calculations, the unit is GeV. }
\end{table}

\section{Conclusion}
In this article, we extend our previous works on the mass spectrum of the
 heavy and doubly heavy baryon states to study the  ${1\over 2}^{\pm}$ and ${3\over 2}^{\pm}$ triply heavy
baryon states   by subtracting the
contributions from the corresponding ${1\over 2}^{\mp}$ and ${3\over 2}^{\mp}$ triply heavy
baryon states with the QCD sum rules, and make reasonable
predictions for their masses.  The  predictions  can be
confronted with the experimental data in the future at the  LHC or used   as   basic input parameters
in other theoretical studies.

\section*{Acknowledgements}
This  work is supported by National Natural Science Foundation,
Grant Number 11075053,  and the Fundamental
Research Funds for the Central Universities.

\section*{Appendix}
The spectral densities of the triply heavy baryon states
 at the level of quark-gluon degrees of freedom,
  \begin{eqnarray}
 \rho^{A}(p_0)&=&\int_{z_{i}}^{z_{f}}dz \int_{y_{i}}^{y_{f}} dy \rho^{A}(p_0,y,z) \, , \nonumber\\
 \rho^{B}(p_0)&=&\int_{z_{i}}^{z_{f}}dz \int_{y_{i}}^{y_{f}} dy \rho^{B}(p_0,y,z) \, ,
 \end{eqnarray}
where
\begin{eqnarray}
\rho^{A,{\frac{1}{2}}^+}_{QQQ'}(p_0,y,z)&=&\frac{3p_0}{8 \pi^4}  yz(1-y-z)(p_0^2-\widetilde{m}^2_{QQ'})(5p_0^2-3\widetilde{m}^2_{QQ'})+\frac{3m_Q^2p_0}{8\pi^4} z(p_0^2-\widetilde{m}^2_{QQ'})\nonumber\\
&&-\frac{1}{24\pi^2}\langle\frac{\alpha_sGG}{\pi}\rangle\left[ \frac{y(1-y-z) m_{Q'}^2}{z^2}+\frac{z (1-y-z)m_{Q}^2}{y^2}+\frac{y z m_{Q}^2}{(1-y-z)^2}\right]\nonumber\\
&&\left[1+\frac{p_0}{4T}\right]\delta(p_0-\widetilde{m}_{QQ'})-\frac{m_{Q'}^2m_Q^2}{192\pi^2p_0T}\langle\frac{\alpha_sGG}{\pi}\rangle \frac{1}{z^2}\delta(p_0-\widetilde{m}_{QQ'})\nonumber\\
&&-\frac{m_{Q}^4}{192\pi^2p_0T}\langle\frac{\alpha_sGG}{\pi}\rangle \left[\frac{z}{y^3}+\frac{z}{(1-y-z)^3}\right]\delta(p_0-\widetilde{m}_{QQ'})\nonumber\\
&&+\frac{m_{Q}^2}{32\pi^2}\langle\frac{\alpha_sGG}{\pi}\rangle \left[\frac{z}{y^2}+\frac{z}{(1-y-z)^2}\right]\delta(p_0-\widetilde{m}_{QQ'})\nonumber\\
&&+\frac{p_0}{32\pi^2}\langle\frac{\alpha_sGG}{\pi}\rangle \left[y-(1-y-z)\right]\left[3+\frac{p_0}{2}\delta(p_0-\widetilde{m}_{QQ'})\right]\nonumber\\
&&+\frac{m_Q^2}{64\pi^2}\langle\frac{\alpha_sGG}{\pi}\rangle \left[\frac{1}{1-y-z}-\frac{1}{y}\right] \delta(p_0-\widetilde{m}_{QQ'}) \, ,
\end{eqnarray}

\begin{eqnarray}
\rho^{B,{\frac{1}{2}}^+}_{QQQ'}(p_0,y,z)&=&\frac{3m_{Q'}}{8 \pi^4}  y(1-y-z)(p_0^2-\widetilde{m}^2_{QQ'})(2p_0^2-\widetilde{m}^2_{QQ'})
 +\frac{3m_{Q'}m_Q^2}{4\pi^4}  (p_0^2-\widetilde{m}^2_{QQ'})\nonumber\\
&&-\frac{m_{Q'}^3}{96\pi^2}\langle\frac{\alpha_sGG}{\pi}\rangle  \frac{y(1-y-z)}{z^3}\left[\frac{1}{p_0}+\frac{1}{2T}\right]\delta(p_0-\widetilde{m}_{QQ'})\nonumber\\
&&-\frac{m_{Q'}m_Q^2}{96\pi^2}\langle\frac{\alpha_sGG}{\pi}\rangle  \left[\frac{(1-y-z)}{y^2}+\frac{y }{(1-y-z)^2}\right]\left[\frac{1}{p_0}+\frac{1}{2T}\right]\delta(p_0-\widetilde{m}_{QQ'})\nonumber\\
&&-\frac{m_{Q'}m_Q^2}{96\pi^2p_0^2T}\langle\frac{\alpha_sGG}{\pi}\rangle \left[\frac{m_{Q'}^2}{z^3}+\frac{m_Q^2}{(1-y-z)^3}+\frac{m_Q^2}{y^3}\right]\delta(p_0-\widetilde{m}_{QQ'})\nonumber\\
&&+\frac{m_{Q'}}{8\pi^2}\langle\frac{\alpha_sGG}{\pi}\rangle \frac{y(1-y-z)}{z^2}\left[1+\frac{p_0}{4}\delta(p_0-\widetilde{m}_{QQ'})\right]\nonumber\\
&&+\frac{m_{Q'}m_Q^2}{16\pi^2p_0}\langle\frac{\alpha_sGG}{\pi}\rangle \left[\frac{1}{y^2}+\frac{1}{z^2}+\frac{1}{(1-y-z)^2}\right]\delta(p_0-\widetilde{m}_{QQ'})\nonumber\\
&&-\frac{m_{Q'} }{16\pi^2}\langle\frac{\alpha_sGG}{\pi}\rangle \left[ 1-\frac{y}{z}+\frac{1-y-z}{z}\right]  \left[1+\frac{p_0}{4}\delta(p_0-\widetilde{m}_{QQ'})\right]\nonumber\\
&&+\frac{m_{Q'}m_Q^2 }{32\pi^2p_0}\langle\frac{\alpha_sGG}{\pi}\rangle \frac{1}{z } \left[\frac{1}{1-y-z}-\frac{1}{y }\right] \delta(p_0-\widetilde{m}_{QQ'}) \, ,
\end{eqnarray}

\begin{eqnarray}
\rho^{A,{\frac{3}{2}}^+}_{QQQ'}(p_0,y,z)&=&\frac{3p_0}{16 \pi^4}  yz(1-y-z)(p_0^2-\widetilde{m}^2_{QQ'})(2p_0^2-\widetilde{m}^2_{QQ'})
+\frac{3m_Q^2p_0}{16\pi^4} z(p_0^2-\widetilde{m}^2_{QQ'}) \nonumber\\
&&-\frac{1}{192\pi^2}
\langle\frac{\alpha_sGG}{\pi}\rangle\left[ \frac{y(1-y-z)m_{Q'}^2}{z^2}+\frac{z(1-y-z)m_{Q}^2}{y^2}+\frac{y z m_{Q}^2}{(1-y-z)^2}\right]\nonumber\\
&&\left[1+\frac{p_0}{2T}\right]\delta(p_0-\widetilde{m}_{QQ'})-\frac{m_{Q'}^2m_Q^2}{384\pi^2p_0T}\langle\frac{\alpha_sGG}{\pi}\rangle \frac{1}{z^2} \delta(p_0-\widetilde{m}_{QQ'})\nonumber\\
&&-\frac{m_Q^4}{384\pi^2p_0T}\langle\frac{\alpha_sGG}{\pi}\rangle \left[ \frac{z}{y^3} +\frac{z}{(1-y-z)^3}\right]\delta(p_0-\widetilde{m}_{QQ'})\nonumber\\
&&+\frac{m_Q^2}{64\pi^2}\langle\frac{\alpha_sGG}{\pi}\rangle \left[ \frac{z}{y^2} +\frac{z}{(1-y-z)^2}\right] \delta(p_0-\widetilde{m}_{QQ'})\nonumber\\
&&-\frac{p_0}{48\pi^2}\langle\frac{\alpha_sGG}{\pi}\rangle z\left[1+\frac{p_0}{8}\delta(p_0-\widetilde{m}_{QQ'})\right]\, ,
\end{eqnarray}

\begin{eqnarray}
\rho^{B,{\frac{3}{2}}^+}_{QQQ'}(p_0,y,z)&=&\frac{3m_{Q'}}{32 \pi^4} y(1-y-z)(p_0^2-\widetilde{m}^2_{QQ'})(3p_0^2-\widetilde{m}^2_{QQ'})
 +\frac{3m_{Q'}m_Q^2}{16\pi^4}  (p_0^2-\widetilde{m}^2_{QQ'})\nonumber\\
&&-\frac{m_{Q'}^3}{384\pi^2T}\langle\frac{\alpha_sGG}{\pi}\rangle  \frac{y(1-y-z)}{z^3} \delta(p_0-\widetilde{m}_{QQ'})\nonumber\\
&&-\frac{m_{Q'}m_Q^2}{384\pi^2p_0^2T}\langle\frac{\alpha_sGG}{\pi}\rangle \left[ \frac{m_{Q'}^2}{z^3}+\frac{m_Q^2}{y^3}+\frac{m_Q^2}{(1-y-z)^3}\right] \delta(p_0-\widetilde{m}_{QQ'})\nonumber\\
&&+\frac{m_{Q'}}{32\pi^2}\langle\frac{\alpha_sGG}{\pi}\rangle  \frac{y(1-y-z)}{z^2} \left[1+\frac{p_0}{2}\delta(p_0-\widetilde{m}_{QQ'})\right]\nonumber\\
&&-\frac{m_{Q'}m_Q^2}{384\pi^2T}\langle\frac{\alpha_sGG}{\pi}\rangle \left[ \frac{y}{(1-y-z)^2}+\frac{1-y-z}{y^2}\right] \delta(p_0-\widetilde{m}_{QQ'})\nonumber\\
&&+\frac{m_{Q'}m_Q^2}{64\pi^2p_0}\langle\frac{\alpha_sGG}{\pi}\rangle  \left[\frac{1}{y^2}+\frac{1}{(1-y-z)^2}+\frac{1}{z^2} \right]\delta(p_0-\widetilde{m}_{QQ'})\nonumber\\
&&+\frac{m_{Q'}}{192\pi^2}
\langle\frac{\alpha_sGG}{\pi}\rangle \left[1+\frac{p_0}{2}\delta(p_0-\widetilde{m}_{QQ'})\right]\, ,
\end{eqnarray}

\begin{eqnarray}
\rho^{A,{\frac{1}{2}}^-}_{QQQ'}(p_0,y,z)&=&\rho^{A,{\frac{1}{2}}^+}_{QQQ'}(p_0,y,z)\mid_{m_{Q'}\to -m_{Q'}}\, , \nonumber\\
\rho^{B,{\frac{1}{2}}^-}_{QQQ'}(p_0,y,z)&=&\rho^{B,{\frac{1}{2}}^+}_{QQQ'}(p_0,y,z)\mid_{m_{Q'}\to -m_{Q'}}\, , \nonumber\\
\rho^{A,{\frac{3}{2}}^-}_{QQQ'}(p_0,y,z)&=&\rho^{A,{\frac{3}{2}}^+}_{QQQ'}(p_0,y,z)\mid_{m_{Q'}\to -m_{Q'}}\, , \nonumber\\
\rho^{B,{\frac{3}{2}}^-}_{QQQ'}(p_0,y,z)&=&\rho^{B,{\frac{3}{2}}^+}_{QQQ'}(p_0,y,z)\mid_{m_{Q'}\to -m_{Q'}}\, ,
\end{eqnarray}
where $z_{f}=\frac{p_0^2+m_{Q'}^2-4m_Q^2+\sqrt{(p_0^2+m_{Q'}^2-4m_Q^2)^2-4p_0^2m_{Q'}^2} }{2p_0^2}$, $z_{i}=\frac{p_0^2+m_{Q'}^2-4m_Q^2-\sqrt{(p_0^2+m_{Q'}^2-4m_Q^2)^2-4p_0^2m_{Q'}^2} }{2p_0^2}$,
$y_f=\frac{1-z+\sqrt{(1-z)^2-\frac{4z(1-z)m_Q^2}{zp_0^2-m_{Q'}^2}}}{2}$, and
$y_i=\frac{1-z-\sqrt{(1-z)^2-\frac{4z(1-z)m_Q^2}{zp_0^2-m_{Q'}^2}}}{2}$. We can take the limit $m_{Q'}=m_{Q}$, and
 obtain the corresponding QCD spectral densities of
the triply heavy baryon states $QQQ$.

\end{document}